\documentclass[11pt,dvips]{article}
\usepackage{eurosym}
\usepackage{graphicx}
\usepackage{amsmath}
\usepackage{amssymb}
\usepackage{latexsym}
\usepackage{color}
\usepackage{epstopdf}
\usepackage[symbol]{footmisc}

\input{tcilatex}
\begin{document}

\thispagestyle{empty}

\begin{center}
\vspace{0.7cm}

{\Large \textbf{Manipulating the direction of one-way steering in an
optomechanical ring cavity}} 

\vspace{0.3cm} \textbf{Jamal El Qars}$^{1}$\textbf{\footnote[1]{%
j.elqars@uiz.ac.ma}, Benachir Boukhris}$^{1,2}$\textbf{, Ahmed Tirbiyine}$%
^{1}$\textbf{, and Abdelaziz Labrag}$^{1}$

$^{1}$\textit{LMS3E, Faculty of Applied Sciences, Ibn Zohr University,
Agadir, Morocco}

$^{2}$\textit{LASIME, National School of Applied Sciences, Ibn Zohr
University, Agadir, Morocco}

\vspace{0.9cm}\textbf{Abstract}
\end{center}

Quantum steering refers to the apparent possibility of exploiting
nonseparable quantum correlations to remotely influence the quantum state of
an observer via local measurements. Different from entanglement and Bell
nonlocality, quantum steering exhibits an inherent asymmetric property,
which makes it relevant for many asymmetric quantum information processing
tasks. Here, we study Gaussian quantum steering between two mechanical modes
in an optomechanical ring cavity. Using experimentally feasible parameters,
we show that the state of the two considered modes can exhibit two-way
steering and even one-way steering. Instead of using unbalanced losses or
noises, we propose a simple practical way to control the direction of
one-way steering. A comparative study between the steering and entanglement
of the studied modes shows that both steering and entanglement undergo a 
\textit{sudden death}-like behavior. In particular, steering is found more
fragile against thermal noise remaining constantly upper bounded by
entanglement. The proposed scheme may be meaningful for one-sided
device-independent quantum key distribution, where the security of such
protocol depends fundamentally on the direction of steering. 
\newline

\bigskip \textbf{Keywords: optomechanical system, covariance matrix,
entanglement, steering, Gaussian states.}

\section{Introduction}

Einstein, Podolsky, and Rosen (EPR) in their seminal 1935 paper \cite{EPR},
have highlighted that when two spatially disjoint particles share an
entangled state, a local measurement implemented on one particle engenders
an apparent nonlocal impact on the second. For them, such effect is
incompatible with the completeness of quantum mechanics. To explain this
phenomenon, commonly known as the EPR paradox, Schr\"{o}dinger introduced
the notion of steering as a nonlocal action that allows the preparation of
quantum states by performing local measurements \cite{Sch1,Sch2}.

The concept of steering has been rigorously defined in terms of violations
of local hidden state model \cite{WJD} as an intermediate form of
nonseparable quantum correlations that stands between entanglement \cite%
{Horo} and Bell nonlocality \cite{Bell}. In a bipartite state $\rho _{%
\mathcal{AB}}$, the violation of Bell-inequality implies the steering in
both directions $\mathcal{A}\rightleftarrows \mathcal{B}$, while steering in
any direction implies that the parties $\mathcal{A}$ and $\mathcal{B}$ are
entangled \cite{WJD}. Passing from entanglement to Bell nonlocality demands
less number of observers and apparatuses that should be trusted \cite{WJD}.
However, their corresponding protocols are proven to be more and more
fragile against decoherence effect \cite{Saunders}.

In quantum information theory, quantum steering can be viewed as an
entanglement verification operation \cite{WJD}, where it certifies the
presence of entanglement in a bipartite quantum state assuming trusted
measurements only on one party. More precisely, if Alice and Bob share a
steerable state $\rho _{\mathcal{AB}}$ at least in one direction (e.g., from
Alice to Bob), then, Alice can convince Bob who does not trust her that the
shared state $\rho _{\mathcal{AB}}$ is entangled by performing local
operations and classical communication \cite{uola}.

Based on the Heisenberg's principle, an experimental criterion for detecting
the EPR paradox was proposed in \cite{Reid1}. Importantly, it has been shown
that violation of such criterion under Gaussian measurements demonstrates
EPR steering \cite{WJD}. The first experimental verification of steering was
realised in \cite{ou}, and followed by other experiments \cite%
{Saunders,wittmann,smith}.

Nowadays, the detection of EPR steering can be accomplished with the help of
various inequalities \cite{Steering inq,wollmann}, where their violation
confirms the existence of EPR steering \cite{Brunner}, but cannot quantify
it \cite{Kogias2}. Then, to quantify the amount by which a bipartite state
is steerable in a given direction, miscellaneous steering measures were
introduced, e.g., the steering weight \cite{Skrzypczyk}, the steering
robustness \cite{Watrous} as well as the Gaussian quantum steering \cite%
{Kogias1}.

Unlike entanglement and Bell non-locality, quantum steering is asymmetric 
\cite{WJD}, i.e., a bipartite state $\rho _{\mathcal{AB}}$ may be steerable,
saying, from $\mathcal{A}\rightarrow \mathcal{B}$, but not vice versa. It is
now believed that the key ingredient of secure quantum information protocols
is asymmetric steering \cite{uola}, which has been recognized as the
essential resource for quantum secret sharing \cite{qsc}, one-way quantum
computing \cite{owqc}, one-sided device-independent quantum key distribution
(1SDI-QKD) \cite{1sDI}, secure quantum teleportation \cite{Reid's},
subchannel discrimination \cite{Watrous}, and other related protocols \cite%
{uola}.

EPR steering has been investigated theoretically as well as experimentally
in various systems \cite%
{Handchen,Armstrong,Bowles,qubit,magnons,elqars1,elqars2,PRD,cluster}. Here,
in an optomechanical ring cavity coupled to a single-mode squeezed light and
driven in the red sideband, we investigate Gaussian quantum steering between
two non-interacting mechanical modes labelled as $\mathcal{A}$ and $\mathcal{%
B}$. Also, we compare the steering of the modes $\mathcal{A}$ and $\mathcal{B%
}$ with their entanglement quantified by the logarithmic negativity \cite%
{vidal,LN}. We show, using realistic experimental parameters, that strong
asymmetric steering, and even one-way steering, can be generated between the
two considered modes. Different from most methods that use unbalanced losses
or noises to control the direction of steering, here we show that the
one-way steering directivity can be practically controlled on demand by
adjusting the coupling between the two mechanical modes $\mathcal{A}$ and $%
\mathcal{B}$ and the cavity field. This therefore provides a flexible and
feasible way in experiments.

In the past decades, much attention has been paid to cavity optomechanics 
\cite{Aspelmeyer} as a potential platform to test different quantum effects.
Proposals include cooling of a mechanical oscillator to its ground state 
\cite{Cooling}, quantum squeezing \cite{Liao}, macroscopic superposition
state \cite{Marshall}, quantum state transfer \cite{transfer}, and quantum
entanglement \cite{entanglement}.

This paper is organized as follows. In \textbf{Section }\ref{sec2}, we
introduce the optomechanical system under consideration. Next, on the basis
of the quantum Langevin equation, we obtain in the resolved sideband regime
the steady-state covariance matrix describing the whole system. In \textbf{%
Section }\ref{sec3}, we quantify and study the Gaussian quantum steering of
two spatially separated mechanical modes and their entanglement. In \textbf{%
Section }\ref{sec4}, we draw our conclusions.

\section{The model and its dynamics \label{sec2}}

\begin{figure}[tbh]
\centerline{\includegraphics[width=8.5cm]{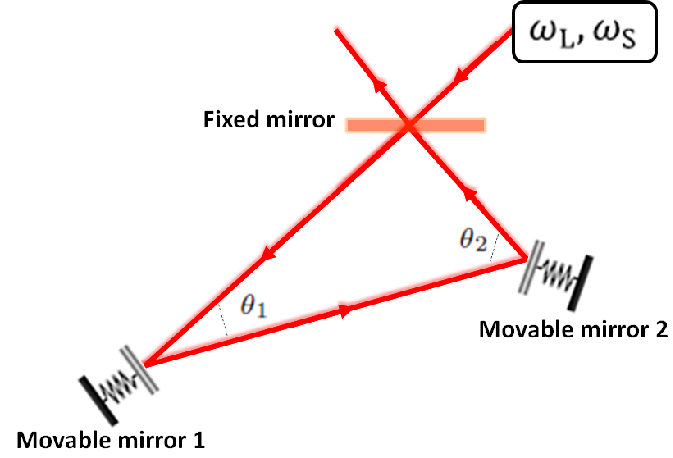}}
\caption{An optomechanical ring cavity driven, through the partially
transmitting fixed mirror, by a coherent laser filed of frequency $\protect%
\omega_{\text{L}}$, and fed by a single-mode squeezed light of frequency $%
\protect\omega_{\text{S}}$.}
\label{f1}
\end{figure}

The system under investigation (see Fig. \ref{f1}), is an optomechanical
ring cavity with one fixed partially transmitting mirror and two movable
perfectly reflecting mirrors. The ring cavity is driven by a coherent laser
with power ${\wp }$, phase $\varphi $ and frequency $\omega _{\text{L}}$.
Also, the ring cavity is fed by a single-mode squeezed light of frequency $%
\omega _{\text{S}}$. The $j\text{th}$ movable mirror is modeled as a
single-mode quantum harmonic oscillator with annihilation operator $b_{j}$,
an effective mass $m_{j}$, frequency $\omega _{m_{j}}$, and damping rate $%
\gamma _{j}$.

\noindent

The Hamiltonian of the system can be written as \cite{Agarwal} 
\begin{equation}
\hat{\mathcal{H}}=\hbar \omega _{c}a^{\dag }a+\sum_{j=1}^{2}\hbar \omega
_{m_{j}}b_{j}^{\dag }b_{j}+\sum_{j=1}^{2}(-1)^{j+1}\hbar g_{_{j}}\cos ^{2}(%
\frac{\theta _{j}}{2})a^{\dag }a(b_{j}^{\dag }+b_{j})+\hbar \varepsilon
(a^{\dag }e^{\mathrm{i}\varphi }+ae^{-\mathrm{i}\varphi }),  \label{e1}
\end{equation}%
where the first term is the free Hamiltonian of the cavity mode with
annihilation operator $a$, frequency $\omega _{c}$ and decay rate $\kappa $.
The second term is the free Hamiltonian of the two mechanical modes. The
third term represents the optomechanical coupling, via the radiation
pressure effect, between the $j$\textrm{th} mechanical mode and the cavity
mode, with coupling strength $g_{_{j}}=\frac{\omega _{c}}{l_{j}}\sqrt{\frac{%
\hbar }{m_{_{j}}\omega _{m_{j}}}}$, $l_{j}$ being the distance between the $%
j $\textrm{th} movable mirror and the fixed mirror. The last therm traduces
the coupling between the input laser and the intracavity mode with coupling
strength $\varepsilon =\sqrt{\frac{\kappa {\wp }}{\hbar \omega _{\text{L}}}}$%
. 

Since the studied system is a dissipative-noisy optomechanical system, then,
its dynamics can be fully described in the Heisenberg picture by means of
the quantum Langevin equation \cite{grebogi}, i.e., $\partial _{t}\mathcal{O}%
=\frac{1}{\mathrm{i}\hbar }\left[ \mathcal{O},\mathcal{\hat{H}}\right]
+dissipation$ $and$ $noise$ $terms$ ($\mathcal{O\equiv }a,b_{j}$). Hence, in
a frame rotating with the laser frequency $\omega _{\text{L}}$, one gets 
\begin{eqnarray}
\dot{b}_{j} &=&-\left( \frac{\gamma _{j}}{2}+\mathrm{i}\omega
_{m_{j}}\right) b_{j}+(-1)^{j}\mathrm{i}g_{_{j}}\cos ^{2}(\frac{\theta _{j}}{%
2})a^{\dag }a+\sqrt{\gamma _{j}}b_{j}^{in},  \label{e2} \\
\dot{a} &=&-\left( \frac{\kappa }{2}-\mathrm{i}\Delta _{0}\right)
a+\sum_{j=1}^{2}(-1)^{j}\mathrm{i}g_{_{j}}\cos ^{2}(\frac{\theta _{j}}{2}%
)a(b_{j}^{\dag }+b_{j})-\mathrm{i}\varepsilon e^{i\varphi }+\sqrt{\kappa }%
a^{in},  \label{e3}
\end{eqnarray}%
where $\Delta _{0}=\omega _{\text{L}}-\omega _{c}$ is the laser detuning 
\cite{Aspelmeyer}. In Eq. (\ref{e2}), $b_{j}^{in}$ denotes the zero-mean
Brownian noise operator affecting the $j\mathrm{th}$ mechanical mode. In
general, this operator is not $\delta $-correlated, characterized by a
non-Markovian correlation function between two instants $t$ and $t^{\prime }$
\cite{Genes}. However, using mechanical resonators with high mechanical
quality factor, i.e., $\mathcal{Q}_{j}=\omega _{m_{j}}/\gamma _{j}\gg 1$,
the Markovian process can be recovered and, consequently, quantum effects
can be achieved. In this limit, we have \cite{Benguria} 
\begin{eqnarray}
\langle b_{j}^{in\dag }(t)b_{j}^{in}(t^{\prime })\rangle &=&n_{\mathrm{th}%
,j}\delta (t-t^{\prime }),\text{ \ }  \label{e4} \\
\langle b_{j}^{in}(t)b_{j}^{in\dag }(t^{\prime })\rangle &=&\left( n_{%
\mathrm{th},j}+1\right) \delta (t-t^{\prime }),\text{ }  \label{e5}
\end{eqnarray}%
where $n_{\mathrm{th},j}=(e^{\hbar \omega _{m_{j}}/k_{B}T_{j}}-1)^{-1}$ is
the mean thermal phonon number in the $j\mathrm{th}$ mechanical mode. $T_{j}$
and $k_{B}$ are, respectively, the temperature of the $j\mathrm{th}$ movable
mirror and the Boltzmann constant. The operator $a^{in}$ that appears in Eq. %
\ref{e3} is the zero mean input squeezed noise operator with the correlation
functions \cite{Agarwal} 
\begin{eqnarray}
\langle a^{in^{\dag }}(t)a^{in}(t^{\prime })\rangle &=&N\delta (t-t^{\prime
}),  \label{e6} \\
\langle a^{in}(t)a^{in^{\dag }}(t^{\prime })\rangle &=&\left( N+1\right)
\delta (t-t^{\prime }),  \label{e7} \\
\langle a^{in}(t)a^{in}(t^{\prime })\rangle &=&Me^{-\mathrm{i}\omega
_{m}\left( t+t^{\prime }\right) }\delta (t-t^{\prime }),  \label{e8} \\
\langle a^{in^{\dag }}(t)a^{in^{\dag }}(t^{\prime })\rangle &=&Me^{\mathrm{i}%
\omega _{m}\left( t+t^{\prime }\right) }\delta (t-t^{\prime }),  \label{e9}
\end{eqnarray}%
where $N=\sinh ^{2}r$ and $M=\sinh r\cosh r$, with $r$ being the squeezing
parameter (we have assumed that $\omega _{m_{1}}=\omega _{m_{2}}=\omega _{m}$%
). 

Due to the quadratic terms $a^{\dag }a$, $ab_{j}^{\dag }$ and $ab_{j}$, Eqs.
(\ref{e2}) and (\ref{e3}) are nonlinear, therefore cannot be solved exactly 
\cite{Genes}. However, by assuming weak optomechanical coupling between the
optical mode $a$ and the $j\mathrm{th}$ mechanical mode $b_{j}$, the
fluctuations $\delta a$ and $\delta b_{j}$ are much smaller than the mean
values $\langle a\rangle $ and $\langle b_{j}\rangle $, then the
linearization of the dynamics can be accomplished around the steady-state.
For this, we write each operator $\mathcal{O}$ as sum of its steady-state
mean value $\langle \mathcal{O}\rangle $ and a small fluctuation $\delta 
\mathcal{O}$ with zero mean value, i.e., $\mathcal{O}=\langle \mathcal{O}%
\rangle +\delta \mathcal{O}$ ($\mathcal{O}\equiv a,b_{j}$) \cite{Milburn}.
Next, by setting $\frac{d}{dt}=0$ in Eqs. (\ref{e2}) and (\ref{e3}) and
factorizing the obtained equations, we obtain $\langle a\rangle =\frac{-%
\mathrm{i}\varepsilon e^{\mathrm{i}\varphi }}{\frac{\kappa }{2}-\mathrm{i}%
\Delta },$ and $\langle b_{j}\rangle =\frac{(-1)^{j}\mathrm{i}g_{_{j}}\cos
^{2}(\frac{\theta _{j}}{2})}{\frac{\gamma _{j}}{2}+\mathrm{i}\omega _{m}}%
\left\vert \langle a\rangle \right\vert ^{2},$ where $\Delta =\Delta
_{0}+\sum_{j=1}^{2}(-1)^{j}g_{_{j}}\cos ^{2}(\frac{\theta _{j}}{2})(\langle
b_{j}\rangle ^{\ast }+\langle b_{j}\rangle )$ denotes the effective detuning 
\cite{Aspelmeyer}. Furthermore, we assume that the cavity is driven by
intense laser, i.e., $\left\vert \langle a\rangle \right\vert \gg 1$ \cite%
{Genes}, then the quadratic fluctuations $\delta a^{\dag }\delta a$, $\delta
a\delta b_{j}$, and $\delta a\delta b_{j}^{\dag }$ can be safely neglected.
Hence, we get 
\begin{eqnarray}
\delta \dot{b}_{j} &=&-\left( \frac{\gamma _{j}}{2}+\mathrm{i}\omega
_{m}\right) \delta b_{j}+(-1)^{j+1}G_{j}\cos ^{2}(\frac{\theta _{j}}{2}%
)\left( \delta a-\delta a^{\dag }\right) +\sqrt{\gamma _{j}}b_{j}^{in},
\label{e10} \\
\delta \dot{a} &=&-\left( \frac{\kappa }{2}-\mathrm{i}\Delta \right) \delta
a+\sum_{j=1}^{2}(-1)^{j}G_{j}\cos ^{2}(\frac{\theta _{j}}{2})\left( \delta
b_{j}+\delta b_{j}^{\dag }\right) +\sqrt{\kappa }a^{in},  \label{e11}
\end{eqnarray}%
where 
\begin{equation}
G_{j}=g_{j}\left\vert \langle a\rangle \right\vert =\sqrt{\frac{\omega
_{c}^{2}\kappa \wp }{l_{j}^{2}m_{j}\omega _{m}\omega _{L}\left( \frac{\kappa
^{2}}{4}+\Delta ^{2}\right) }},  \label{e12}
\end{equation}%
is the effective coupling.

We emphasize that Eqs. (\ref{e10}) and (\ref{e11}) are obtained using $%
\langle a\rangle =-\mathrm{i}\left\vert \langle a\rangle \right\vert $ or
equivalently to $\tan \varphi =-2\Delta /\kappa $. In what follows, we
introduce the operators $\delta \tilde{b}_{j}=\delta b_{j}e^{\mathrm{i}%
\omega _{m}t}$ and $\delta \tilde{a}=\delta ae^{-\mathrm{i}\Delta t}$ \cite%
{Genes} and we consider that the system is driven in the red sideband ($%
\Delta =-\omega _{m}$), which is convenable for quantum-state transfer from
the squeezed light to the mechanical modes \cite{Aspelmeyer}. In addition,
in the resolved-sideband regime, where $\omega _{m}\gg \kappa $, the
rotating wave approximation allows to neglect terms rotating at $\pm 2\omega
_{m}$ \cite{RWA}. Therefore, we get 
\begin{eqnarray}
\delta \dot{\tilde{b}}_{j} &=&-\frac{\gamma _{j}}{2}\delta \tilde{b}%
_{j}+(-1)^{j+1}G_{j}\cos ^{2}(\frac{\theta _{j}}{2})\delta \tilde{a}+\sqrt{%
\gamma _{j}}b_{j}^{in},  \label{e13} \\
\delta \dot{\tilde{a}} &=&-\frac{\kappa }{2}\delta \tilde{a}%
+\sum_{j=1}^{2}(-1)^{j}G_{j}\cos ^{2}(\frac{\theta _{j}}{2})\delta \tilde{b}%
_{j}+\sqrt{\kappa }a^{in}.  \label{e14}
\end{eqnarray}

Now, using Eqs. (\ref{e13}) and (\ref{e14}) and the quadratures position and
momentum of the $j\mathrm{th}$ mechanical(optical) mode $\delta \tilde{q}%
_{j}=(\delta \tilde{b}_{j}^{\dag }+\delta \tilde{b}_{j})/\sqrt{2}$ and $%
\delta \tilde{p}_{j}=\mathrm{i}(\delta \tilde{b}_{j}^{\dag }-\delta \tilde{b}%
_{j})/\sqrt{2}$ $\Big(\delta \tilde{x}=(\delta \tilde{a}^{\dag }+\delta 
\tilde{a})/\sqrt{2}$ and $\delta \tilde{y}=\mathrm{i}(\delta \tilde{a}^{\dag
}-\delta \tilde{a})/\sqrt{2}\Big)$ as well as the $j\mathrm{th}$ input
mechanical(optical) noise operators $\tilde{q}_{j}^{in}=(b_{j}^{in\dagger
}+b_{j}^{in})/\sqrt{2}$ and $\tilde{p}_{j}^{in}=\mathrm{i}(b_{j}^{in\dagger
}-b_{j}^{in})/\sqrt{2}$ $\Big(\tilde{x}^{in}=(a^{in\dag }+a^{in})/\sqrt{2}$
and $\tilde{y}^{in}=\mathrm{i}(a^{in\dag }-a^{in})/\sqrt{2}\Big)$, we obtain

\begin{eqnarray}
\partial _{t}\delta \tilde{q}_{j} &=&-\frac{\gamma _{j}}{2}\delta \tilde{q}%
_{j}+(-1)^{j+1}G_{j}\cos ^{2}(\frac{\theta _{j}}{2})\delta \tilde{x}+\sqrt{%
\gamma _{j}}\tilde{q}_{j}^{in},  \label{e15} \\
\partial _{t}\delta \tilde{p}_{j} &=&-\frac{\gamma _{j}}{2}\delta \tilde{p}%
_{j}+(-1)^{j+1}G_{j}\cos ^{2}(\frac{\theta _{j}}{2})\delta \tilde{y}+\sqrt{%
\gamma _{j}}\tilde{p}_{j}^{in},  \label{e16} \\
\partial _{t}\delta \tilde{x} &=&\sum_{j=1}^{2}(-1)^{j}G_{j}\cos ^{2}(\frac{%
\theta _{j}}{2})\delta \tilde{q}_{j}-\frac{\kappa }{2}\delta \tilde{x}+\sqrt{%
\kappa }\tilde{x}^{in},  \label{e17} \\
\partial _{t}\delta \tilde{y} &=&\sum_{j=1}^{2}(-1)^{j}G_{j}\cos ^{2}(\frac{%
\theta _{j}}{2})\delta \tilde{p}_{j}-\frac{\kappa }{2}\delta \tilde{y}+\sqrt{%
\kappa }\tilde{y}^{in},  \label{e18}
\end{eqnarray}%
which can be written as $\partial _{t}\tilde{u}=\mathcal{A}\tilde{u}+\tilde{n%
}$, with $\tilde{u}^{\mathrm{T}}=(\delta \tilde{q}_{1},\delta \tilde{p}%
_{1},\delta \tilde{q}_{2},\delta \tilde{p}_{2},\delta \tilde{x},\delta 
\tilde{y})$, $\tilde{n}^{\mathrm{T}}=(\tilde{q}_{1}^{in},\tilde{p}_{1}^{in},%
\tilde{q}_{2}^{in},\tilde{p}_{2}^{in},\tilde{x}^{in},\tilde{y}^{in})$ and 
\begin{equation}
\mathcal{A}=\left( 
\begin{array}{cccccc}
\frac{-\gamma _{1}}{2} & 0 & 0 & 0 & G_{1}\cos ^{2}(\frac{\theta _{1}}{2}) & 
0 \\ 
0 & \frac{-\gamma _{1}}{2} & 0 & 0 & 0 & G_{1}\cos ^{2}(\frac{\theta _{1}}{2}%
) \\ 
0 & 0 & \frac{-\gamma _{2}}{2} & 0 & -G_{2}\cos ^{2}(\frac{\theta _{2}}{2})
& 0 \\ 
0 & 0 & 0 & \frac{-\gamma _{2}}{2} & 0 & -G_{2}\cos ^{2}(\frac{\theta _{2}}{2%
}) \\ 
-G_{1}\cos ^{2}(\frac{\theta _{1}}{2}) & 0 & G_{2}\cos ^{2}(\frac{\theta _{2}%
}{2}) & 0 & \frac{-\kappa }{2} & 0 \\ 
0 & -G_{1}\cos ^{2}(\frac{\theta _{1}}{2}) & 0 & G_{2}\cos ^{2}(\frac{\theta
_{2}}{2}) & 0 & \frac{-\kappa }{2}%
\end{array}%
\right) .  \label{e19}
\end{equation}

Because the dynamics of the system is linearized and the operators $%
b_{j}^{in}$ and $a^{in}$ are zero-mean quantum Gaussian noises, the steady
state of the quantum fluctuations is a zero-mean tripartite Gaussian state,
fully described by its $6\times 6$ covariance matrix $\mathcal{V}$ defined
by $\mathcal{V}_{kk^{\prime }}=\langle \tilde{u}_{k}(\infty )\tilde{u}%
_{k^{\prime }}(\infty )+\tilde{u}_{k^{\prime }}(\infty )\tilde{u}_{k}(\infty
)\rangle /2$ \cite{Genes}.

Using standard approach \cite{vitali}, one can determine the matrix $%
\mathcal{V}$ by solving the following Lyapunov equation 
\begin{equation}
\mathcal{A}\mathcal{V}+\mathcal{V}\mathcal{A}^{\mathrm{T}}=-\mathcal{D},
\label{e20}
\end{equation}%
where the diffusion matrix $\mathcal{D}$ is defined by $\mathcal{D}=\langle 
\tilde{n}_{j}(t)\tilde{n}_{j^{\prime }}(t^{\prime })+\tilde{n}_{j^{\prime
}}(t^{\prime })\tilde{n}_{j}(t)\rangle /2=\mathcal{D}_{jj^{\prime }}\delta
(t-t^{\prime })$ \cite{vitali}, and found to be $\mathcal{D}=\frac{\gamma
_{1}}{2}\left( 2n_{\mathrm{th,1}}+1\right) 
\mbox{$1
\hspace{-1.0mm}  {\bf l}$}_{2}\oplus \frac{\gamma _{2}}{2}\left( 2n_{\mathrm{%
th,2}}+1\right) \mbox{$1
\hspace{-1.0mm}  {\bf l}$}_{2}\oplus \mathrm{diag}\left( \frac{\kappa }{2}%
e^{2r},\frac{\kappa }{2}e^{-2r}\right) .$

Equation \ref{e20} is a linear equation for $\mathcal{V}$ and can be
straightforwardly solved, however, the exact expression of $\mathcal{V}$ is
too cumbersome and can not be reported here. The covariance matrix $\mathcal{%
V}_{m}$ of the mechanical modes $\mathcal{A}$ and $\mathcal{B}$ can be
obtained by tracing over the optical mode variables in the matrix $\mathcal{V%
}$. Then, we have 
\begin{equation}
\mathcal{V}_{m}=\left( 
\begin{array}{cc}
\mathcal{V}_{\mathcal{A}} & \mathcal{V}_{\mathcal{A/B}} \\ 
\mathcal{V}_{\mathcal{A/B}}^{\mathrm{T}} & \mathcal{V}_{\mathcal{B}}%
\end{array}%
\right) ,  \label{e21}
\end{equation}%
where the $2\times 2$ block matrices $\mathcal{V}_{\mathcal{A}}$ and $%
\mathcal{V}_{\mathcal{B}}$ represent the mechanical modes $\mathcal{A}$ and $%
\mathcal{B}$, respectively. While $\mathcal{V}_{\mathcal{A/B}}$ describes
the correlations between them.

\section{Gaussian steering of the two mechanical modes\label{sec3}}

A generic two-mode Gaussian state $\rho _{\mathcal{AB}}$ with covariance
matrix $\mathcal{V}_{m}$ given by Eq. (\ref{e21}), is steerable, under
Gaussian measurements \cite{gs} performed on mode $\mathcal{A}$, if the
following condition is violated \cite{WJD,Kogias1} 
\begin{equation}
\mathcal{V}_{m}+\mathrm{i}(0_{\mathcal{A}}\oplus \Omega _{\mathcal{B}%
})\geqslant 0,  \label{e22}
\end{equation}%
where $0_{\mathcal{A}}$ is a $2\times 2$ null matrix and $\Omega _{\mathcal{B%
}}$ $=\left( 
\begin{array}{cc}
0 & 1 \\ 
-1 & 0%
\end{array}%
\right) $ is the $\mathcal{B}$-mode symplectic matrix \cite{Kogias1}.
Quantitatively, the amount by which the state $\rho _{\mathcal{AB}}$ is
steerable, under Gaussian measurements performed on mode $\mathcal{A}$, is
given by \cite{Kogias1} 
\begin{equation}
\mathcal{G}^{\mathcal{A\rightarrow B}}=\max \left[ 0,\frac{1}{2}\ln \frac{%
\det \mathcal{V}_{\mathcal{A}}}{4\det \mathcal{V}_{m}}\right] ,  \label{e23}
\end{equation}%
where $\mathcal{G}^{\mathcal{B\rightarrow A}}$ can be obtained by changing
the roles of $\mathcal{A}$ and $\mathcal{B}$ in Eq. (\ref{e23}), i.e., $%
\mathcal{G}^{\mathcal{B\rightarrow A}}=\max \left[ 0,\frac{1}{2}\ln \frac{%
\det \mathcal{V}_{\mathcal{B}}}{4\det \mathcal{V}_{m}}\right] $.

Since quantum steering is asymmetric, then the state $\rho _{\mathcal{AB}}$
may be steerable in one direction, but not vice versa \cite{Kogias1}. In
general, we distinguish: (\textit{i}) no-way steering, where the state $%
\rho_{\mathcal{AB}}$ is nonsteerable in any direction, i.e., $\mathcal{G}^{%
\mathcal{A\rightarrow B}}=\mathcal{G}^{\mathcal{B\rightarrow A}}=0$, (%
\textit{ii}) two-way steering, where the state $\rho_{\mathcal{AB}}$ is
steerable in both directions, i.e., $\mathcal{G}^{\mathcal{A\rightarrow B}%
}>0 $ and $\mathcal{G}^{\mathcal{B\rightarrow A}}>0$, (\textit{iii}) one-way
steering where the state $\rho_{\mathcal{AB}}$ is steerable solely in one
direction, i.e., $\mathcal{G}^{\mathcal{A\rightarrow B}}>0$ with $\mathcal{G}%
^{\mathcal{B\rightarrow A}}=0 $ or $\mathcal{G}^{\mathcal{B\rightarrow A}}>0$
with $\mathcal{G}^{\mathcal{A\rightarrow B}}=0 $.

Besides, to compare the behavior of the steering of the two mechanical modes 
$\mathcal{A}$ and $\mathcal{B}$ with their entanglement, we use the
logarithmic negativity $\mathcal{E}_{\mathcal{N}}$ defined by $\mathcal{E}_{%
\mathcal{N}}=\max \left[ 0,-\ln 2\nu \right] $ \cite{vidal,LN}, where $\nu =%
\sqrt{\left( \sigma -\sqrt{\sigma ^{2}-4\det V_{m}}\right) /2}$ is the
smallest simplistic eigenvalue of the partial transpose of the $4\times 4$
covariance matrix $V_{m}$, with $\sigma =\det V_{\mathcal{A}}+\det V_{%
\mathcal{B}}-2\det V_{\mathcal{A/B}}$. We note that the logarithmic
negativity $\mathcal{E}_{\mathcal{N}}$ is necessary and sufficient
entanglement monotone for Gaussian states \cite{vidal,LN}.

For achieving asymmetric steering, it is necessary to introduce asymmetry
into the system. In this respect, we consider the case where $G_{1}\neq
G_{2} $, which can practically be realized by choosing identical parameters
in Eq. (\ref{e12}), except the lengths $l_{1}$ and $l_{2}$. Moreover, to
have fairly good idea of steering and entanglement of the modes $\mathcal{A}$
and $\mathcal{B}$, we use realistic experimental parameters from \cite%
{Groblacher,Arcizet}. The movable mirrors have a mass $m_{1,2}=m=145~\mathrm{%
ng}$, oscillate at frequency $\omega _{m_{1,2}}=\omega _{m}=2\pi \times 947~%
\mathrm{KHz}$, and damped at rate $\gamma _{1,2}=\gamma =2\pi \times 140~%
\mathrm{Hz}$. The ring cavity, characterized by a decay rate $\kappa =2\pi
\times 215~\mathrm{KHz}$ and frequency $\omega _{c}=2\pi \times 5.26\times
10^{14}~\mathrm{Hz}$, is pumped by a coherent laser field of frequency $%
\omega _{\text{L}}=2\pi \times 2.82\times 10^{14}$ $\mathrm{Hz}$ and power $%
\wp =50~\text{mW}$. Here, it is interesting to notice that since the system
is driven within the red sideband, the stability conditions are always
satisfied regardless of the chosen parameters \cite{Genes}.

\begin{figure}[t]
\centerline{\includegraphics[width=0.45\columnwidth,height=5cm]{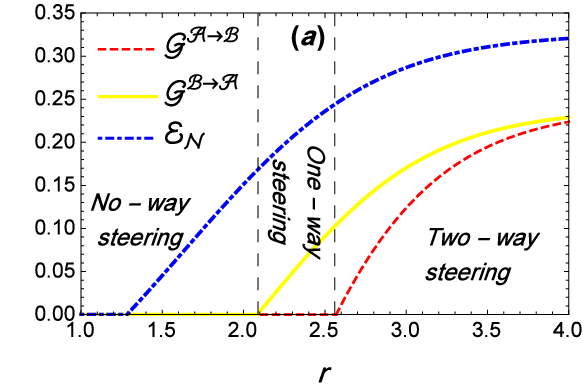}
\includegraphics[width=0.45\columnwidth,height=5cm]{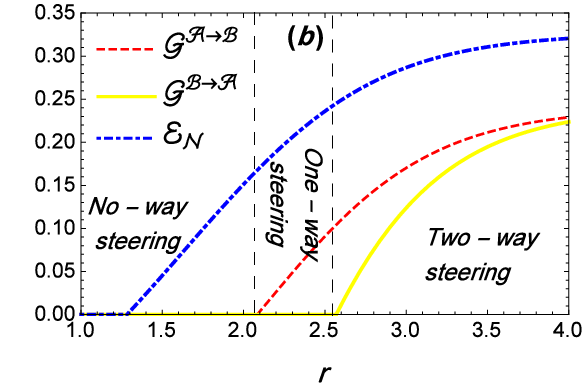}}
\caption{The steerabilities $\mathcal{G}^{\mathcal{A\rightarrow B}}$ (red
dashed line) and $\mathcal{G}^{\mathcal{B\rightarrow A}}$ (yellow solid
line), and entanglement $\mathcal{E}_{\mathcal{N}}$ (blue dot-dashed line)
of the modes $\mathcal{A}$ and $\mathcal{B}$ versus the squeezing parameter $%
r$. In (a) we used $l_{1}=112~\protect\mu \mathrm{m}$, $l_{2}=85~\protect\mu 
\mathrm{m}$, $\protect\theta _{1}=\protect\pi /6$ and $\protect\theta _{2}=%
\protect\pi /3$, where $l_{j}$ is the distance between the fixed mirror and
the $j\text{th}$ movable mirror. In (b) we used $l_{1}=85~\protect\mu 
\mathrm{m}$, $l_{2}=112~\protect\mu \mathrm{m}$, $\protect\theta _{1}=%
\protect\pi /3$ and $\protect\theta _{2}=\protect\pi /6$. In both panels, we
used $n_{\text{th},1}=n_{\text{th},2}=5$ as value of the mean thermal phonon
numbers. Obviously, in panel (a) where $l_{1}/l_{2}>1$, we remark that
one-way steering is occurred from $\mathcal{B\rightarrow A}$, while in panel
(b) where $l_{1}/l_{2}<1$ one-way steering is occurred in the reverse
direction $\mathcal{A\rightarrow B}$. Then one concludes that the direction
of one-way steering could be practically controlled via the lengths $l_{1}$
and $l_{2}$. }
\label{f2}
\end{figure}

Figure \ref{f2} shows the squeezing influence on the steerabilities $%
\mathcal{G}^{\mathcal{A}\rightarrow \mathcal{B}}$ and $\mathcal{G}^{\mathcal{%
B}\rightarrow \mathcal{A}}$, and entanglement $\mathcal{E}_{\mathcal{N}}$ of
the two mechanical modes $\mathcal{A}$ and $\mathcal{B}$. In Fig. \ref{f2}%
(a), we used $l_{1}=112~\mu \text{m}$, $l_{2}=85~\mu \text{m}$, $\theta
_{1}=\pi /6$, and $\theta _{2}=\pi /3$. While in Fig. \ref{f2}(b), we used $%
l_{1}=85~\mu \text{m}$, $l_{2}=112~\mu \text{m}$, $\theta _{1}=\pi /3$, and $%
\theta _{2}=\pi /6$. For the mean thermal phonon numbers, we used $n_{\text{%
th},1}=n_{\text{th},2}=5$ in both cases. As shown, the steerable states are
always entangled, but, entangled ones are not always steerable. On the other
hand, Fig. \ref{f2} reveals that, in comparison with entanglement, steering
requires high values of squeezing $r$ to be created, meaning that steering
is a form of nonseparable quantum correlations stronger than entanglement.
We emphasize that squeezed light with squeezing parameter $r>3$ is well
attained in \cite{Vahlbruch}.

Manifestly, the steerabilities $\mathcal{G}^{\mathcal{A}\rightarrow \mathcal{%
B}}$ and $\mathcal{G}^{\mathcal{B}\rightarrow \mathcal{A}}$, and
entanglement $\mathcal{E}_{\mathcal{N}}$ behave in the same way under
influence of the squeezing parameter $r$. This can be explained as follows:
the progressive injection of squeezed light increases the number of photons
in the cavity, which in turn acts positively on the optomechanical coupling
strengths $G_{1}$ and $G_{2}$. This therefore enhances entanglement and
steering between the modes $\mathcal{A}$ and $\mathcal{B}$, where they
increase gradually to their maximum. Moreover, Figs. \ref{f2}(a) and \ref{f2}%
(b) show that by interchanging the values of the lengths $l_{1}$ and $l_{2}$%
, and then the values of the coupling strengths $G_{1}$ and $G_{2}$,
entanglement $\mathcal{E}_{\mathcal{N}}$ is not sensitive to such operation,
in contrast, the steering $\mathcal{G}^{\mathcal{A\rightarrow B}}$ and $%
\mathcal{G}^{\mathcal{B\rightarrow A}}$ are strongly influenced.

Quite remarkably, $\mathcal{G}^{\mathcal{A\rightarrow B}}$ and $\mathcal{G}^{%
\mathcal{B\rightarrow A}}$ are asymmetric, where two-way steering and even
one-way steering can be displayed by the state $\rho _{\mathcal{AB}}$ as can
be seen from Fig. \ref{f2}. For example, Fig. \ref{f2}(a) shows that for $%
2\leq r\leq 2.5$, the state $\rho _{\mathcal{AB}}$ is one-way steerable from 
$\mathcal{B\rightarrow A}$ where $\mathcal{G}^{\mathcal{A\rightarrow B}}=0$
and $\mathcal{G}^{\mathcal{B\rightarrow A}}>0$. While for $r>2.5$, the state 
$\rho _{\mathcal{AB}}$ is two-way steerable, where both $\mathcal{G}^{%
\mathcal{A\rightarrow B}}$ and $\mathcal{G}^{\mathcal{B\rightarrow A}}$ are
strictly positive. Fig. \ref{f2}(a) together with Fig. \ref{f2}(b) show that
for $r\geq 2.5$, it is possible to obtain two-way steering behavior over a
wide range of the squeezing parameter $r$, which has been proven to be a
necessary resource needed for teleporting a coherent state with fidelity
beyond the nocloning limit \cite{Reid's}. The different degree of steering
observed in Figs. \ref{f2}(a) and \ref{f2}(b) between the directions $%
\mathcal{A\rightarrow B}$ and $\mathcal{B\rightarrow A}$ is also shown to
provide the asymmetric guaranteed key rate achievable within a practical
1SDI-QKD \cite{Kogias1}.

Importantly, Fig. \ref{f2} shows that the direction of one-way steering
between the modes $\mathcal{A}$ and $\mathcal{B}$ could be practically
controlled on demand via the lengths $l_{1}$ and $l_{2}$, where $l_{j}$ is
the distance between the $j\text{th}$ movable mirror and the fixed mirror.
Indeed, in Fig. \ref{f2}(a), where we used $l_{1}=112~\mu \text{m}$ and $%
l_{2}=85~\mu \text{m}$, one-way steering is occurred from $\mathcal{%
B\rightarrow A}$. In contrast, for $l_{1}=85~\mu \text{m}$ and $%
l_{2}=112~\mu \text{m}$, one-way steering is occurred in the reverse
direction $\mathcal{A\rightarrow B}$ as illustrated in Fig. \ref{f2}(b).
This therefore provides a flexible and feasible experimental way for
manipulating the direction of one-way steering, which offers a reference for
the practical application of EPR steering.

One-way steering observed, for example, from $\mathcal{B\rightarrow A}$
could be interpreted as follows: Alice(owning mode $\mathcal{A}$) and
Bob(owning mode $\mathcal{B}$) can implement the same Gaussian measurements
on their shared state $\rho _{\mathcal{AB}}$, however, obtain different
results. In addition, Bob can convince Alice that their shared state is
entangled, while Alice cannot. Such asymmetric behavior is partly related to
the asymmetry introduced between the two modes $\mathcal{A}$ and $\mathcal{B}
$, and partly related to the definition of the aspect of steering in terms
of the EPR paradox \cite{Reid1,Kogias1}. The most obvious application of
one-way steering property is that it provides security in 1SDI-QKD protocol,
where the measurement device of one party only is untrusted. 
\begin{figure}[t]
\centerline{\includegraphics[width=0.45\columnwidth,height=5cm]{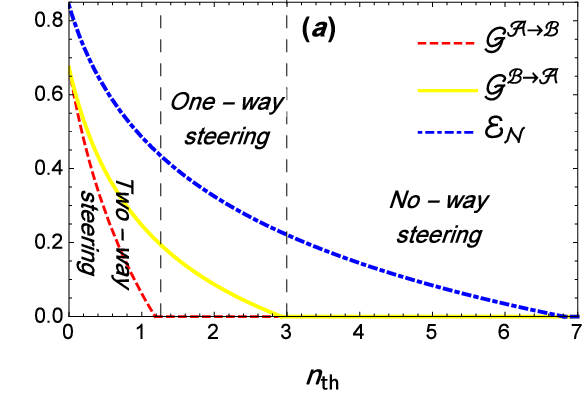}
\includegraphics[width=0.45\columnwidth,height=5cm]{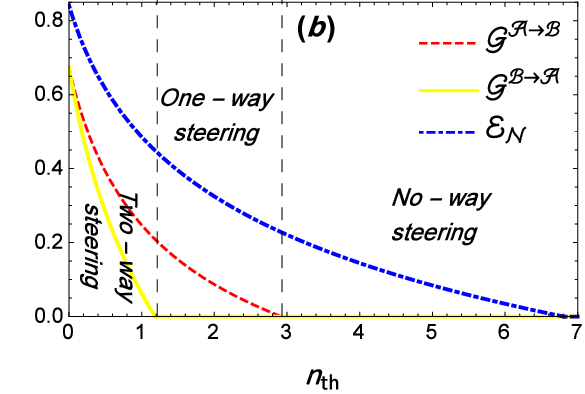}}
\caption{The steerabilities $\mathcal{G}^{\mathcal{A\rightarrow B}}$ (red
dashed line) and $\mathcal{G}^{\mathcal{B\rightarrow A}}$ (yellow solid
line), and entanglement $\mathcal{E}_{\mathcal{N}}$ (blue dot-dashed line)
of the two modes $\mathcal{A}$ and $\mathcal{B}$ versus the common mean
thermal phonon number $n_{\text{th},1}=n_{\text{th},2}=n_{\text{th}}$. The
lengths $l_{1}$ and $l_{2}$, and the angles $\protect\theta _{1}$ and $%
\protect\theta _{2}$ are the same as in Fig. \protect\ref{f2}. In both
panels, we used $r=1.5$ as value of the squeezing parameter.}
\label{f3}
\end{figure}

In Fig. \ref{f3} we plot $\mathcal{G}^{\mathcal{\mathcal{A}\rightarrow B}}$, 
$\mathcal{G}^{\mathcal{B\rightarrow A}}$ and $\mathcal{E}_{\mathcal{N}}$
versus the common mean thermal phonon number $n_{\mathrm{th},1}=n_{\mathrm{th%
},2}=n_{\mathrm{th}}$ using $r=1.5$ for the squeezing parameter. The lengths 
$l_{1}$, $l_{2}$ and the angles $\theta _{1}$, $\theta _{2}$ are the same as
in Figs. \ref{f2}(a) and \ref{f2}(b). As can be seen, the three measures $%
\mathcal{G}^{\mathcal{A\rightarrow B}}$, $\mathcal{G}^{\mathcal{B\rightarrow
A}}$ and $\mathcal{E}_{\mathcal{N}}$ are maximum for $n_{\text{th}}=0$, and
they decrease with increasing $n_{\text{th}}$. Strikingly, the
steerabilities $\mathcal{G}^{\mathcal{A\rightarrow B}}$ and $\mathcal{G}^{%
\mathcal{B\rightarrow A}}$ remain constantly upper bounded by entanglement $%
\mathcal{E}_{\mathcal{N}}$, and have a tendency to decay rapidly, than
entanglement, to zero under thermal noise. This indicates that quantum
steering is more fragile than entanglement against decoherence effect.
Similarly to the results depicted in Fig. \ref{f2}, Fig. \ref{f3} shows that
the steerable states are always entangled, but entangled ones are not in
general steerable. This means that nonzero degree of entanglement is
indispensable for steering.

More interestingly, Fig. \ref{f3} shows that thermal noise, not only
deteriorates the degree of entanglement and steering, but can play a
positive role in realizing and orienting one-way steering by the mediation
of the ratio $l_{1}/l_{2}$. Indeed, in Fig. \ref{f3}(a), where $%
l_{1}/l_{2}>1 $, we remark that by increasing $n_{\text{th}}$, the
steerability $\mathcal{G}^{\mathcal{A\rightarrow B}}$ vanishes for $n_{\text{%
th}}\geq 1.2$, besides this $\mathcal{G}^{\mathcal{B\rightarrow A}}$ still
persists and can be detected for $n_{\text{th}}=3$, which means that the
state $\rho _{\mathcal{AB}}$ is one-way steerable from $\mathcal{%
B\rightarrow A}$ for $1.2<n_{\text{th}}<3$. Whereas, in Fig. \ref{f3}(b),
where $l_{1}/l_{2}<1$, it is clearly seen that by increasing $n_{\text{th}}$%
, the steering $\mathcal{G}^{\mathcal{B\rightarrow A}}$ becomes zero for $n_{%
\text{th}}\geq 1.2$, in contrast $\mathcal{G}^{\mathcal{A\rightarrow B}}$
can be detected until $n_{\text{th}}=3$, meaning that the state $\rho _{%
\mathcal{AB}}$ is one-way steerable in the direction $\mathcal{A\rightarrow B%
}$ for $1.2<n_{\text{th}}<3$.

Our scheme shows an important advantage, where one-way steering can be
generated and manipulated without imposing asymmetric losses or noises
between the considered modes $\mathcal{A}$ and $\mathcal{B}$, but
controlling the optomechanical coupling strengths $G_{1}$ and $G_{2}$
through the distances $l_{1}$ and $l_{2}$, where $l_{j}$ is the distance
between the $j$\textrm{th} movable mirror and the fixed mirror (see Fig. \ref%
{f1}). This may offer a precious resource for 1SDI-QKD, knowing that the
security of such protocol depends crucially on the direction of steering 
\cite{1sDI}.

Finally, on the basis of the strategy developed in \cite{vitali}, it is
possible to swap the quantum correlations from the two mechanical modes $%
\mathcal{A}$ and $\mathcal{B}$ back to two auxiliary optical modes.
Furthermore, employing a single homodyne detector technique \cite{shd}, the
entire correlation matrix of the two optical auxiliary modes can be
reconstructed, which in turn allows us to determine the elements of the
matrix $V_{m}$ given by Eq. (\ref{e21}) and from them one can numerically
estimate $\mathcal{G}^{\mathcal{A\rightarrow B}}$ and $\mathcal{G}^{\mathcal{%
B\rightarrow A}}$.

\section{Conclusions \label{sec4}}

In an optomechanical ring cavity fed by squeezed light and driven in the red
sideband, stationary Gaussian quantum steering of two mechanical modes $%
\mathcal{A}$ and $\mathcal{B}$ is studied. In the resolved sideband regime,
the steady-state covariance matrix describing the two considered modes is
obtained. Using realistic experimental parameters, we showed that stronger
asymmetric steering can be generated between the modes $\mathcal{A}$ and $%
\mathcal{B}$. Also, we showed that the two-mode Gaussian state $\rho _{%
\mathcal{AB}}$ can exhibit one-way steering from $\mathcal{A}\rightarrow 
\mathcal{B}$ as well as from $\mathcal{B}\rightarrow \mathcal{A}$.
Essentially, we showed that the direction of one-way steering could be
practically controlled via the lengths $l_{1}$ and $l_{2}$, where $l_{j}$ is
the distance between the $j\text{th}$ movable mirror and the fixed mirror
(see Fig. \ref{f1}). This therefore offers a flexible way for controlling
the direction of one-way steering in experimental operations. The change of
one-way steering direction observed in Figs. \ref{f2} and \ref{f3} may lead
to the change of the role played by two communication parties, which is
shown to play a decisive role in establishing more security in 1SDI-QKD
protocol \cite{1sDI}.

Besides, a comparison study between the steering of the two modes $\mathcal{A%
}$ and $\mathcal{B}$ with their entanglement quantified by means of the
logarithmic negativity $\mathcal{E}_{\mathcal{N}}$ showed on the one hand
that steerable states are always entanglement, but entangled state are not
necessarily steerable, on the other hand both steering and entanglement
undergo a \textit{sudden death}-like behavior under thermal effect. In
particular, steering is found more fragile, than entanglement, against
thermal effect, decays rapidly to zero, and remains upper bounded by the
degree of entanglement.

Our one-way steering manipulation scheme is facile to be accomplished
experimentally, which provides a reference for the practical application of
asymmetric Gaussian quantum steering based on robust entangled states.

\end{document}